\documentclass{icrc2009}

\usepackage{graphicx}   % for including figures
\usepackage{fixltx2e}
\usepackage{url}
\newcommand{\shorttitle}[1]%
{\markboth{Proceedings of the 31\MakeLowercase{$^{st}$} ICRC, {\L}\'{o}d\'{z} 2009}{#1} }
\newcommand{\etal}{\MakeLowercase{\textit{et al. }}} % "et al."

\begin{document}
\title{Search for Exotic Physics with the ANTARES Detector}

\author{\IEEEauthorblockN{%St\'ephanie Escoffier\IEEEauthorrefmark{1},
			  Gabriela Pavalas\IEEEauthorrefmark{1} and
                          Nicolas Picot Clemente\IEEEauthorrefmark{2},
                          %Vlad Popa\IEEEauthorrefmark{2} 
                          \\
			  on behalf of the ANTARES Collaboration}
                            \\
\IEEEauthorblockA{\IEEEauthorrefmark{1}Institute for Space Sciences, Bucharest-Magurele, Romania}
\IEEEauthorblockA{\IEEEauthorrefmark{2}CNRS / Centre de Physique des Particules de Marseille, Marseille, France}}

% please write the preseter's name and short title (3-4 words maximum)
%    which will appear at the header of the even pages.
\shorttitle{G. Pavalas \etal Exotic Physics with ANTARES}
\maketitle

\begin{abstract}
%Very rare exotic objects that could be present in cosmic radiation are currently searched for with neutrino telescopes. 
Besides the detection of high energy neutrinos, the ANTARES telescope offers an opportunity
to improve sensitivity to exotic cosmological relics.  
In this article we discuss the sensitivity of the ANTARES detector to relativistic monopoles and slow nuclearites. 
Dedicated trigger algorithms and search strategies are being developed to search for them. The data filtering, background rejection selection criteria are described, as well as the expected sensitivity of ANTARES
to exotic physics.

\end{abstract}
\section{Introduction}
The ANTARES neutrino telescope is aimed to observe high energy cosmic neutrinos through
the detection of the Cherenkov light produced by up-going induced muons. However, the
ANTARES detector is also sensitive to a variety of exotic particles, and can provide an
unique facility for the search of magnetic monopoles and nuclearites.

\section{The ANTARES detector}

The ANTARES detector has reached its nominal size in May 2008. The 884 Optical Modules (OM) are deployed on
twelve vertical lines in the Western Mediterranean, at depths between 2050 and 2400 meters.
The OMs, consisting of a glass sphere housing a 10'' Hamamatsu photomultiplier (PMT)~\cite{Ant_PMT}, are arranged by triplet per storey. Each detector line, made of 25 storeys, is connected via interlinks to a Junction Box, itself connected to the 
shore station at La Seyne-sur-Mer through a 40 km long electro-optical cable. The strategy 
of the ANTARES data acquisition is based on the ``all-data-to-shore'' concept~\cite{Ant_Acq}.
This implementation leads to the transmission of all raw data above a given threshold to shore, where different
triggers are applied to the data for their filtering before storage. 
\\
For the analysis presented here, only the two general trigger logics operated up to now have been considered. 
Both are based on local coincidences. A local coincidence (L1 hit) is defined either 
as a combination of two hits on two OMs of the same storey within 20 ns, or as a single hit 
with a large amplitude, typically 3 pe. The first trigger, a so-called directional trigger, 
requires five local coincidences anywhere in the detector but causally connected, within a 
time window of 2.2 $\mu$s. The second trigger, a so-called cluster trigger, requires two
T3-clusters within 2.2 $\mu$s, where a T3-cluster is a combination of two L1 hits in adjacent 
or next-to-adjacent storeys. When an event is triggered, all PMT pulses are recorded over
2.2 $\mu$s.
\\
The ANTARES observatory was build gradually, 
giving rise to various detector layouts used for physics analysis. The 5-line, 10-line and 
12-line detector configurations match with data taken from January 2007, from December 2007 
and from May 2008, respectively. 

\section{Magnetic monopoles}
\subsection{Introduction}

Most of the Grand Unified Theories (GUTs) predict the creation of magnetic monopoles in the early Universe. Indeed, in 1974, 't Hooft \cite{thooft} and Polyakov \cite{polyakov} showed independently that each time a compact and connected gauge group is broken into a connected subgroup, elements caracterising well a magnetic charge, as introduced by Dirac in 1931 \cite{Dirac}, appear.\\
These particles are topologically stable and carry a magnetic charge defined as a multiple integer of the Dirac charge $g_D=\frac{\hbar c}{2e}$, where e is the elementary electric charge, c the speed of light in vacuum and $\hbar$ the Planck constant. Depending on the group, the masses inferred for magnetic monopoles can take range over many orders of magnitude, from $10^8$ to $10^{17}$ GeV.

As magnetic monopoles are stable, and so would survive until now, they should have been very diluted in the Universe, as predicted by numerous theoretical studies which set stringent limit on their fluxes, like the Parker flux limit \cite{Parker}. More stringent limits were set recently by different experiments (MACRO \cite{MACRO}, AMANDA \cite{AMANDA}, ...).

The development of neutrino astronomy in the last decade led to the construction of huge detectors, which allow new hopes in the search for magnetic monopoles.
Actually, the ANTARES detector seems to be well designed to detect magnetic monopoles, or at least to improve limits on their fluxes, as described below.
\subsection{Signal and background simulations}

Since fast monopoles have a large interaction with matter, they can lose large amounts of energy in the terrestrial environment. The total energy loss of a relativistic monopole with one Dirac charge is of the order of $10^{11}$ GeV \cite{EnergyLoss} after having crossed the full diameter of the Earth. Because magnetic monopoles are expected to be accelerated in the galactic coherent magnetic field domain to energies of about $10^{15}$ GeV \cite{CosmicFlux}, some could be able to cross the Earth and reach the ANTARES detector as upgoing signals.

\begin{figure}[!h]
  \centering
  \includegraphics[width=2.5in]{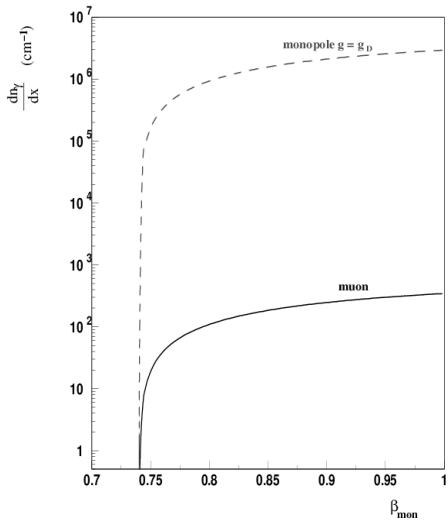}
  \caption{Number of emitted photons per unit length ($cm^{-1}$) by a magnetic monopole with a charge $g=g_D$ through direct Cherenkov emission (dashed line) compared to the number of photons emitted by a muon (black line), as a function of their velocities.}
  \label{nberofphotons}
 \end{figure}

The monopole's magnetic charge $g=n g_D$ can be expressed as an equivalent electric charge $ g = 68.5 n e $, where n is an integer. Thus relativistic monopoles with $\beta\geq0.74$ carrying one Dirac charge will emit a large amount of direct Cherenkov light when traveling through the ANTARES detector, giving rise to $\sim 8500$ more intense light than a muon. The number of photons per unit length ($cm^{-1}$) emitted on the path of a monopole is shown in figure \ref{nberofphotons}, as a function of the velocity of the monopole up to $\gamma = 10 $ ($\beta=0.995$).
 
For the analysis, monopoles have been simulated inside an optimized volume containing the 12-line detector, for six ranges of velocities between $\beta=0.74$ and $\beta=0.995$.
In addition, downgoing atmospheric muons have been simulated using the CORSIKA package \cite{Corsika}, as well as upgoing and downgoing atmospheric neutrinos according to the Bartol flux \cite{Bartol1,Bartol2}. Optical background from $^{40}K$ decay has been added to both magnetic monopole signal and atmospheric muon and neutrino background events.
\subsection{Search strategy}

The 12-line detector data are triggered by both trigger logics, the directional and the cluster triggers (see section II). A comparison of efficiency was therefore performed on magnetic monopoles and restricted only to upward-going events. 
As the efficiency of the directional trigger was found to be lower than for the cluster trigger, it was decided to perform searches for upgoing magnetic monopoles with the cluster trigger only.

\begin{figure}[!h]
  \centering
  \includegraphics[width=2.5in]{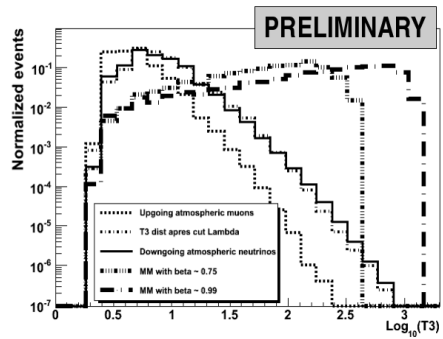}
  \caption{Normalized events as a function of the number of T3 clusters for downgoing atmospheric muons, upgoing and downgoing atmospheric neutrinos, and upgoing magnetic monopoles with $\beta \sim 0.75$ and $\beta \sim 0.99$.}
  \label{DistribT3Norm}
 \end{figure}

The standard reconstruction algorithm, developed in ANTARES for upward-going neutrino selection, and mainly based on a likelihood maximization, was applied. In order to select upgoing particles, only reconstructed events with a zenith angle lower than 90$^{\circ}$ were selected.
However, muon bundles are difficult to reconstruct properly and some of them can be reconstructed as upward-going events.\\
As it is shown in figure \ref{DistribT3Norm}, a magnetic monopole traversing the detector will emit an impressive quantity of light, compared to atmospheric muons or muons induced by atmospheric neutrinos.
The large amount of induced hits in the detector, more precisely the number of T3 clusters, is therefore used as a criteria to remove part of the atmospheric background.

\begin{figure}[!h]
  \centering
  \includegraphics[width=2.5in]{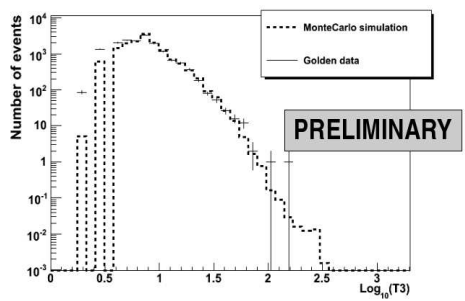}
  \caption{Comparison of T3 distributions between data and Monte Carlo simulations for 10 days of data taking.}
  \label{DistribT3_10days}
 \end{figure}

Before applying a supplementary cut to reduce the remaining background, 10 active days of golden\footnote{Golden data assumes experimental data complying with certain selection criteria like low baserate and burstfraction.} data were taken as reference to check the data Monte Carlo agreement. The comparison of T3 distributions between the data and the background simulation for 10 days is shown in figure \ref{DistribT3_10days}.

We optimized the cuts on the number of T3 clusters to maximize the 90$\%$ C.L. sensitivity, calculated with the usual Feldman-Cousins formula\cite{Feld}, for magnetic monopoles after one year of data taking.
In the optimization process the same selection criteria have been applied to calculate the sensitivity to magnetic monopole events over the whole velocity range $0.74 \leq \beta \leq 0.995$.

Finally the 90$\%$ C.L. sensitivity for this range was found, for a cut of at least 140 T3 clusters, for which about 1.7 background events are expected.
The 90$\%$ C.L. sensitivity for ANTARES after one year of data taking is of the order of $\sim 1\cdot10^{-17}cm^{-2}s^{-1}sr^{-1}$.

\section{Nuclearites}
\subsection{Introduction}
    Nuclearites are hypothetical nuggets of strange quark matter that could be present in
cosmic radiation. Their origin is related to energetic astrophysical phenomena
(supernovae, collapsing binary strange stars, etc.). Down-going nuclearites could reach
the ANTARES depth with velocities $\sim$ 300 km/s, emitting blackbody radiation at visible
wavelengths while traversing sea water.

    Heavy nuggets of strange quark matter ($M\geq10^{10}$GeV), known as nuclearites, would be electrically neutral; the small positive electric charge of the quark core would be neutralized by electrons forming an electronic cloud or found in week equilibrium inside the core. The relevant energy loss mechanism is represented by the elastic collisions with the atoms of the traversed media, as shown in ref. \cite{ruj}:
  \begin{equation}
    \frac{dE}{dx} = -\sigma\rho{v^2},
    \label{en_loss}
   \end{equation}  
where $\rho$ is the density of the medium, $v $ is the nuclearite velocity and $\sigma$ its geometrical cross section:
 
   \begin{displaymath}
        \sigma = \left\{ \begin{array}{ll}
                        \pi(3M/4\pi\rho_N)^{2/3} & \mbox{for $M\geq8.4*10^{14}$ GeV};\\
                       \pi\times10^{-16} \mbox{cm}^2 &  \mbox{for lower masses}.
                    
	             \end{array}
	             \right. 
	\label{sigma_cross}
     \end{displaymath}
with $\rho_N=3.6\times10^{14}$ g cm$^{-3}$. The mass limit in the above equation corresponds to a radius of the strange quark matter of 1\AA. Assuming a nuclearite of mass $M$ enters the atmosphere with an initial (non-relativistic) velocity $v_0$, after crossing a depth L it will be slowed down to

 \begin{equation}
    v(L) = v_0e^{-\frac{\sigma}{M}\int_0^L\rho dx}   
    \label{velocity}
   \end{equation} 
where $\rho$ is the air density at different depths. Considering the parametrization of the standard atmosphere from \cite{shi}:

 \begin{equation}
    \rho(h) = ae^{-\frac{h}{b}}=ae^{-\frac{H-l}{b}}, 
    \label{density}
   \end{equation} 
where a=$1.2\times10^{-3}$ g cm$^{-3}$ and b$\simeq8.57\times10^5$ cm, H$\simeq$50 km is the total height of the atmosphere, the integral in Eq.\ref{velocity} may be solved analytically.

\begin{figure}
\centering
\includegraphics[width=2.5in]{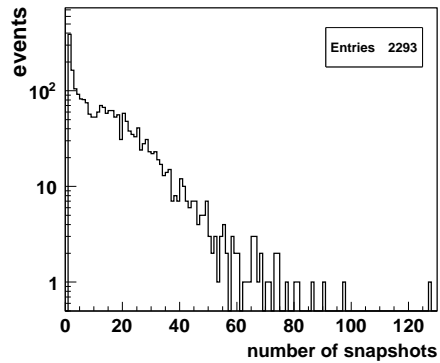}
\caption{Snapshot distribution obtained for simulated nuclearite events with masses 3$\times10^{16}$, 10$^{17}$ and 10$^{18}$ GeV.}
\label{mult}
\end{figure}

The propagation of nuclearites in sea water is described also by Eq. \ref{velocity}, assuming $\rho=1$ g cm$^{-3}$ and substituting $v_0$ with the speed value at the Earth surface. Nuclearites moving into the water could be detected because of the black-body radiation emitted by the expanding cylindrical thermal shock wave \cite{ruj}. The luminous efficiency (defined as the fraction of dissipated energy appearing as light) was estimated, in the case of water, to be $\eta\simeq3\times10^{-5}$ \cite{ruj}. The number of visible photons emitted per unit path length can be computed as follows:
 \begin{equation}
    \frac{dN_{\gamma}}{dx}=\eta\frac{dE/dx}{\pi(eV)},   
    \label{en_eta}
   \end{equation} 
assuming the average energy of visible photons $\pi$ eV.

\subsection{Search strategy and results}

 The Monte Carlo simulation of nuclearite detection in ANTARES assumes only the down-going part of an isotropic flux of nuclearites and an initial velocity (before entering the atmosphere) of $\beta=10^{-3}$. A typical nuclearite event would cross the Antares detector in a characteristic time of $\sim$ 1 ms, producing a luminosity that would exceed that of muons by several orders of magnitude. Down-going atmospheric muons represent the main background for nuclearite events.  
This analysis refers to simulated nuclearite and muon events using the 5-line detector configuration and 84 days of data taken from June to November 2007.   
\\
	Simulated nuclearite and muon events have been processed with the directional trigger, that operated during the 5-line data acquisition. Background was added from a run taken in July 2007, at a baserate of 63.5 kHz. We found for nuclearites a lower mass limit detectable with the directional trigger of $3\times10^{16}$ GeV. 
Nuclearite events were simulated for masses of $3\times10^{16}$ GeV, $10^{17}$ GeV and $10^{18}$ GeV.
The atmospheric muons were generated with the MUPAGE code \cite{MUP}. The parameters used in our analysis comprised the number of L1 triggered hits (see Section II), the number of single hits (L0 hits, defined as hits with a threshold greater than 0.3 photoelectrons), the duration of the snapshot (defined as the time difference between the last and the first L1 triggered hits of the event) and total amplitude of hits in the event. Data were reprocesed with the directional trigger.
We obtained a good agreement between simulated muon events and data.

The algorithm of the directional trigger selects from all the hits produced by a nuclearite only those that comply to the signal of a relativistic muon. These hits can be contained in a single snapshot or multiple snaphots for a single event. 

\begin{figure}
\centering
\includegraphics[width=2.5in]{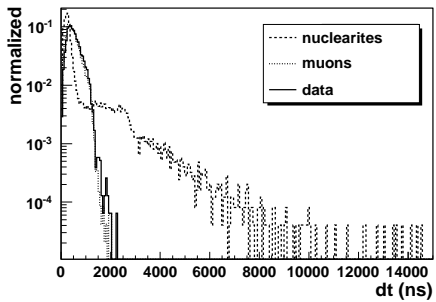}
\caption{Normalized distributions as a function of the duration of snapshot. Comparison between data (continuous line), simulated muon (dotted line) and nuclearite (dashed line) events is shown.}
\label{dt}
\end{figure}

Because nuclearites are slowly moving particles, multiple snapshots belonging to a single nuclearite event may span into intervals from tens of $\mu$s up to $\sim$1ms. The multiplicity of snapshots in the simulated nuclearite events is presented in Fig. \ref{mult}.     
The majority of the snapshots produced by nuclearites from the studied sample are of short duration ($<$500 ns), see Fig. \ref{dt}.  
Long duration snapshots (up to tens of $\mu$s) are also presented, due to the large light output of events with masses $\geq 10^{17}$ Gev.

\begin{figure}
\centering
\includegraphics[width=2.5in]{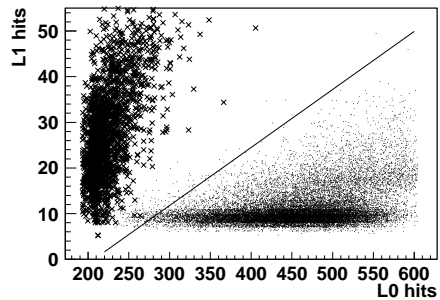}
\caption{Data (crosses) and simulated nuclearite events (points) in a L1 triggered hits vs. L0 single hits plot. The points represent the corresponding values per snapshot. The linear cut was optimized for signal selection.}
\label{L10_hits}
\end{figure}

Snapshots of nuclearites passing through the detector would also be characterised by a larger number of single hits than for atmospheric muons.  
In the following, the selection criteria for nuclearite signal were obtained considering only the data sample as background.
The distributions for data and simulated nuclearite events in the 2-dimensional plot L1 triggered hits vs L0 single hits show a clear separation, that was optimized using the linear cut presented in Fig. \ref{L10_hits}. This cut reduces the data by 99.998$\%$.  

A second cut has been applied to the remaining events, by requiring the multiple snapshot signature within a time interval of 1 ms, characteristic to the crossing time of the detector by a nuclearite event. Three "events" with a double snapshot have passed the second cut. 

The percentage of simulated nuclearite events remained after applying the cuts is given in Table \ref{table_nucl}.
The events below the linear cut are characterized by snapshots with a low number of L1 hits and a large number of single hits.

\begin{table}[!h]
  \caption{Percentage of nuclearite events after cuts}
  \label{table_nucl}
  \centering
  \begin{tabular}{|c|c|c|}
  \hline
   Nuclearite mass (GeV)  &  linear cut & multiple snapshot cut \\
   \hline 
    3$\times10^{16}$ & 72.5\% & 16.3\% \\
    1$\times10^{17}$ & 96.8\% & 89.1\% \\
    1$\times10^{18}$ & 98.7\% & 96.2\% \\
  \hline
  \end{tabular}
  \end{table} 
 
The sensitivity of the ANTARES detector in 5-line configuration after 84 days of data taking to nuclearite events with masses $\geq 10^{17}$ GeV is of the order of $\sim 10^{-16}cm^{-2}s^{-1}sr^{-1}$. 
We are currently investigating the possibility of implementation in the data acquisition program of a trigger that uses the characteristics of nuclearite events in order to keep all data sent to shore for a time interval of about 20 ms. Also, test runs on simulated nuclearite data using the cluster trigger show a better efficiency and a lower detectable mass limit than for directional trigger.  

\section{Conclusions}
We presented search strategies and expected sensitivities for monopoles and non-relativistic nuclearites with the ANTARES detector in 12-line and 5-line configurations. The sensitivities obtained for both upward magnetic monopoles and nuclearites are preliminary. The sensitivity for nuclearites can be improved by using data taken in nominal configuration and the cluster trigger.

\end{document}